\documentclass[amsmath,superscriptaddress,aps,prb,preprintnumbers]{revtex4}
\usepackage{bm}
\usepackage{graphics}
\begin{document}
\pagestyle{empty}

\title{Gapless Surfaces in Anisotropic Superfluids}

\author{E. Gubankova}
\affiliation{Center for Theoretical Physics, Department of
Physics, MIT, Cambridge, Massachusetts 02139}

\author{E.G. Mishchenko}
\affiliation{Departmet of Physics, University of Utah, Salt Lake City, UT 84112}

\author{F. Wilczek}
\affiliation{Center for Theoretical Physics, Department of
Physics, MIT, Cambridge, Massachusetts 02139}

\preprint{MIT-CTP \#3562}
\begin{abstract}
We demonstrate when p-wave pairing occurs between species whose free Fermi surfaces are mismatched the gap generally vanishes
over a two-dimensional surface.  We present detailed calculations of condensation energy, superfluid density (Meissner mass) and specific heat for such states.  We also consider stability against separation into mixed phases.   According to several independent criteria that can be checked at weak coupling, the resulting ``breached'' state appears
to be stable over a substantial range of parameters.    The simple models we consider are homogeneous in position space, and break rotation symmetry spontaneously.  They should be realizable in cold atom systems.    
\end{abstract}

\maketitle

\section{Introduction}

Recently there has been considerable
interest in a class of possible new states featuring coexistence
of superfluid and normal components. These states arise when there
are interactions favoring pairing between fermions that have fermi
surfaces of different size.  For s-wave superfluidity, which has mostly been considered, 
a breach in the pairing occurs in the pairing of fermions which have momenta
whose magnitudes lie between two values $p_\pm$.  Two separate two-dimensional spherical fermi surfaces, corresponding to gapless modes, open up at $|p| = p_\pm$; while paired fermions with momenta outside 
the breach provide a coexisting superfluid condensate.

The possibility of superfluidity coexisting with gapless states at momenta
that span a two-dimensional surface originally was suggested
by Sarma\cite{Sarma,LW}. For spherically
symmetric (s-wave) interactions, as he considered, a state of this type naturally suggests itself,  
and a pairing solution can be found
\cite{Sarma,LW,GLW,ABR,SH}. The stability of the resulting state
against phase separation \cite{Bedaque} or the appearance of a tachyon
in the gauge field (negative squared Meissner mass)
\cite{WuYip,HuangShov,Nardulli} is delicate, however. It appears to require
some combination of unequal masses, momentum-dependent pairing
interactions, and long-range neutrality
constraints\cite{FGLW,AKR}.

Cold atom systems with mismatched free fermi surfaces naturally arise when the
system contains
different species of atoms, or 
identical atoms with different spin states in an external magnetic
field.  Experimental realization of fermion superfluids in cold atoms
is a major recent development in condensed matter physics
\cite{coldatoms}. 
Manipulation of the parameters of Feshbach resonances 
and various trapping techniques provide control over the form and
the strength of interaction and effective masses (band
structure). In a wider context, having a mixture of bosonic and
fermionic atoms and tuning to different parameter regimes, cold
atom systems open possibilities to explore new exotic phases
\cite{exp}.  

Most experiments to date have exploited s-wave Feshbach resonances, 
but p-wave resonance, accessing a state nearly bound by a dipolar interaction, is also
experimentally realizable\cite{Zhang}.

At strong enough coupling s-wave breached pairing is realized near 
Feshbach resonance, opening one or two spherical Fermi
surfaces\cite{Son,recent},
while it is destroyed by instabilities in Meissner mass and
number susceptibility\cite{recent} at weak coupling. In this case, the p-wave
breached superconductor is a stable ground state as we show in this work.
Even when the s-wave interaction is repulsive, p-wave pairing instability
may arise due to the Kohn-Luttinger effect\cite{Kohn-Luttinger}, in the same way
it may support p-wave pairing in liquid $^3{\rm He}$. In QCD with one flavor,
the order parameter
has total angular momentum one, i.e. $l=1$ or $s=1$, and the preferred phase
exhibits color-spin-locking in the relativistic limit\cite{Schaefer}. 

Here we suggest analyze a model
where the breached pair state arises in the p-wave channel.  We find that in this context it is
quite robustly stable.  It seems
not unreasonable, intuitively, that expanding an existing
(lower-dimensional) locus of zeros into a two-dimensional surface
should be significantly easier than producing, as in s-wave, a whole sphere of gapless
excitations ``from scratch''.   In our context, we will show that pairing of mismatched Fermi surfaces can expand the locus of gapless states from a line to a torus (polar phase) or from two points into two lenticles (planar phase).  We shall show that this phenomenon even occurs
for arbitrarily small coupling and small Fermi surface mismatch, where our mean field approximation should be adequate.
We presented a short account of some of this work previously\cite{shortVersion}.

Anisotropic superfluid states that coexist with gapless modes at
isolated points or lines in momentum space are also well known
\cite{Mineev}. Gapless states also are known to occur in the
presence of magnetic impurities \cite{Abrikosov} and,
theoretically, in states with spontaneous breaking of translation
symmetry \cite{LOFF}, where the gapless states span a
two-dimensional fermi surface. A similar phenomenon was found in two-dimensional 
d-wave superconductors subject to an external magnetic field \cite{yangSondhi}.
Strong coupling between different
bands also may lead to zeros in quasiparticle excitations and
gapless states \cite{Volovik}.

A crucial difference between the model we consider and the
conventional p-wave superfluid system, $^{3}$He, lies in the
dissimilarity of the paired species.  Although there are
two components, there is no approximate quasispin symmetry, and no
analogue of the fully gapped B phase \cite{BW} arises.

\section{Anisotropic p-wave pairing}

We consider a model system with the two species of
fermions having the same Fermi velocity $v_F$, but  different
Fermi momenta $p_F\pm I/v_F$. The effective Hamiltonian is
\begin{equation}
\label{ham} H=\sum_{\bf p} [ \epsilon^A_{ p} a_{\bf p}^\dagger
a_{\bf p}+ \epsilon^B_{p} b_{-\bf p}^\dagger b_{\bf -p} -
\Delta_{\bf p}^{*} a_{\bf p}^\dagger b^\dagger _{-\bf p} +
\Delta_{\bf p} b_{-\bf p} a_{\bf p}]
\end{equation}
with $\epsilon^A_{p}=\xi_p +I,~\epsilon^B_{p}=\xi_p -I$,
$\xi_p=v(p-p_F)$, $\Delta_{\bf p}=\sum_{\bf k} V_{\bf p-k}\langle
a_{\bf k}^\dagger b^\dagger _{-\bf k}\rangle$. Here the attractive interspecies
interaction is $-V_{\bf p-k}$ within the ``Debye'' energy $2\omega_D$
around the Fermi surface ($\omega_D\gg I$), and vanishes at larger energies.
The intraspecies
interaction is assumed to be either repulsive or negligibly small.
For the sake of simplicity, we have taken the gap function $\Delta_{\bf p}$ to be real. Excitations of the Hamiltonian (\ref{ham}) are
gapless, $E_{\bf p}=\pm \sqrt{\xi^2_p+\Delta_{\bf p}^2} +I$,
provided that there are areas on the Fermi surface where
$I>\Delta_{\bf p}$. The gap equation at zero temperature,
\begin{equation}
\Delta_{\bf p}=\frac{1}{2}\sum_{\bf k}V_{\bf p-k}
\frac{\Delta_{\bf k}}{\sqrt{\xi^2_k+\Delta_{\bf k}^2}}~
\theta\left(\sqrt{\xi^2_k+\Delta_{\bf k}^2}-I\right),
\end{equation}
can be simplified by taking the integral over $d\xi_k$,
\begin{equation}
\Delta_{\bf n}=\nu \int \frac{d \Omega_{\bf n'}}{4\pi}
V({\bf n,n'}) \Delta_{\bf n'} \left(\ln{\frac{2\omega_D}{|\Delta_{\bf
n'}|}} +\Theta (I-\Delta_{\bf n'})\ln{\frac{|\Delta_{\bf
n'}|}{I+\sqrt{I^2-\Delta^2_{\bf n'}}} } \right).
\label{gapeq}
\end{equation}
Due to hirarchy of scales, $\omega_D\ll E_F$, the integration is performed
in the narrow window around the Fermi surface (BCS approximation), i.e.
$d^3k\rightarrow \nu\, d\Omega\, d\xi_k$,
where $\nu=1/(2\pi)^3\int d^3k\, \delta (\xi_k)=k_F^2/(2\pi^2v)$ is the density
of states.
In deriving
Eq.~(\ref{gapeq}) we neglected dependence of $V_{\bf p-k}$ on the
absolute values of ${\bf p}$ and ${\bf k}$, this is a good
approximation as long as the ``Debye'' energy is small compared to
the Fermi energy, $\omega_D \ll E_F$.

We concentrate on the first harmonic ($p$-wave pairing) in the
expansion of the interaction potential over the spherical
functions,  $ V({\bf n,n'})=g ({\bf n}\cdot {\bf n'})$, with
$g>0$. The gap equation (\ref{gapeq}) allows two solutions
corresponding to the value of the projection of the angular
momentum  of the Cooper pair onto some axis ${\bf z}$: $m=0$
(polar phase), and $m=\pm 1$ (planar phase). We will now analyze
these two solutions in detail.

\subsection{Polar phase: $\Delta_{\bf n} \sim Y_{10}({\bf n})$.}

Let us look for a solution  in the form $\Delta_{\bf
n}=\Delta({\bf z}\cdot{\bf n})$ where the direction ${\bf z}$
reflects a broken rotational symmetry. In spherical coordinates,
with  ${\bf z}\parallel {\bf n}$ and $\theta=({\bf n},{\bf n'})$
the gap equation becomes,
\begin{equation}
\label{gap_planar0}
 -\frac{2}{\nu g}=\int\limits_0^{\pi/2} d\theta
\sin{\theta}\cos^2{\theta} \ln\left(\frac{\Delta\cos{\theta}}{2\omega_D}\right)
+\int\limits_{\theta^*}^{\pi/2} d\theta \sin{\theta}\cos^2\theta
\ln\left(\frac{I+\sqrt{I^2-\Delta^2 \cos^2{\theta} }}{\Delta
\cos{\theta} } \right).
\end{equation}
where $\theta^*=\arccos{(I/\Delta)}$, for $I<\Delta$, and $\theta^*=0$
for $I>\Delta$.



The solution of Eq.~(\ref{gap_planar0}) is different in two cases
of small and large Fermi momentum mismatches,

\subsubsection{Small mismatch, $I<\Delta$.}

Performing the angle integration in Eq.~(\ref{gap_planar0}), we
obtain
\begin{equation}
-\frac{1}{\nu g}=\frac{1}{3}\ln\left(\frac{\Delta}{2\omega_D}\right)
-\frac{1}{9}
+\frac{1}{6}\left(\frac{I}{\Delta}\right)^3\frac{\pi}{2}.
\label{gap}
\end{equation}
At zero Fermi momentum mismatch, $I=0$, the gap is,
$ \Delta_0/2\omega_D= \exp{\left(\frac{1}{3} -\frac{3}{\nu g}\right)}
\approx 1.40 ~ \exp{\left( -\frac{3}{\nu g} \right)}$. Using new
dimensionless variables $x=I/\Delta_0$ and $y=\Delta/\Delta_0$ in
Eq.~(\ref{gap}) we obtain,
\begin{equation}
\label{gap0}
\ln\frac{1}{y}=\frac{\pi}{4}\left(\frac{x}{y}\right)^3.
\end{equation}
At small $x\ll 1$, the solution of this equation is approximated
by $y\approx 1-{\pi x^3}/{4}$.

\subsubsection{Large mismatch, $I>\Delta$.}

The gap equation now takes the form,
\begin{equation}
\label{gap_planar_1}
 -\frac{2}{\nu g}=\int\limits_0^{\pi/2} d\theta \sin{\theta}\cos^2{\theta}
\ln\left(\frac{I+\sqrt{I^2-\Delta^2 \cos^2{\theta} }}{2\omega_D}\right).
\end{equation}
Performing the angle integration we obtain
\begin{equation}
-\frac{1}{\nu
g}=\frac{1}{3}\ln\left(\frac{I+\sqrt{I^2-\Delta^2}}{2\omega_D}\right)
-\frac{1}{9}
-\frac{1}{6}\left(\frac{I}{\Delta}\right)^3\sqrt{1-\left(\frac{\Delta}{I}
\right)^2}
+\frac{1}{6}\left(\frac{I}{\Delta}\right)^3\arcsin\left(\frac{\Delta}{I}\right),
\label{gapI}
\end{equation}
which in the dimensionless variables $x=I/\Delta_0$,
$y=\Delta/\Delta_0$ is simplified to,
\begin{equation}
\ln\frac{1}{x+\sqrt{x^2-y^2}}=-\frac{1}{2}\left(\frac{x}{y}\right)^2
\sqrt{1-\left(\frac{y}{x}\right)^2}
+\frac{1}{2}\left(\frac{x}{y}\right)^3\arcsin\left(\frac{y}{x}\right).
\label{gap0I}
\end{equation}
At $x=x_c=(1/2)e^{-1/3}$ the gap vanishes according to,
$y
\approx 1.55 (x-x_c)^{1/2}. $ The solution of the gap equation
(\ref{gap0}) and (\ref{gap0I}), consists of two branches shown in
Fig.~1.
\begin{figure}[h]
\label{fig2}
\resizebox{.15\textwidth}{!}{\includegraphics{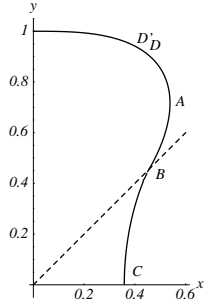}}
\caption{The solutions $y(x)$ of Eqs.~(\ref{gap0},\ref{gap0I}).
The lower branch corresponds to the unstable state. Both branches
merge at $\Delta=I$ (the point B; broken line). Non-zero solutions
of the gap equation cease to exist beyond the point A. Stability
conditions from the energy and the linear response give the points
D and D$^\prime$, respectively. Coordinates of the characteristic points
(for $\Delta_0=1$) are $A((4/3\pi e)^{1/3}=0.538,e^{-1/3}=0.717)$,
$B(e^{-\pi/4}=0.456,0.456)$, $C(0,e^{-1/3}/2=0.358)$,
$D(0.475,0.886)$, $D^\prime(0.440,0.917)$.}
\end{figure}

\subsection{Planar phase: $\Delta_{\bf n} \sim Y_{11}({\bf n})$,
$\Delta_{\bf n} \sim Y_{1-1}({\bf n})$.}

We now look for a solution corresponding to the momentum $m=\pm 1$
of a Cooper pair: $\Delta_{\bf n}=\Delta\sin{({\bf n},{\bf
z})e^{\pm i\phi}}$, here $\phi$ is the polar angle in the plane
perpendicular to ${\bf z}$. The gap equation becomes,
\begin{equation}
\label{gap_planar}
 -\frac{2}{\nu g}=\int\limits_0^{\pi/2} d\theta
\sin^3{\theta} \ln\left(\frac{\Delta\sin{\theta}}{2\omega_D}\right)
+\int\limits_0^{\theta^*} d\theta \sin^3{\theta}
\ln\left(\frac{I+\sqrt{I^2-\Delta^2 \sin^2{\theta} }}{\Delta
\sin{\theta} } \right).
\end{equation}
where $\theta^*=\arcsin{(I/\Delta)}$, for $I<\Delta$, and
$\theta^*=\pi/2$, for $I>\Delta$. Again, we consider separately
the regimes of small and large Fermi momentum mismatches.

\subsubsection{Small mismatch, $I<\Delta$.}

The first integral in Eq.~(\ref{gap_planar}) is simple: $
\int_0^{\pi/2} d\theta \sin^3{\theta}
\ln{[\sin{\theta}]}=\frac{2}{3}\ln{2} -\frac{5}{9}. $ The second
integral in Eq.~(\ref{gap_planar}), denoted by $K_1(I)$, can be
most easily calculated in the following way. Let us first
differentiate $K_1(I)$ with respect to $I$,
\begin{equation}
\label{der_k1} \frac{\partial K_1}{\partial
I}=\int\limits_0^{\arcsin{(I/\Delta)}} d\theta
\frac{\sin^3{\theta}}{ \sqrt{I^2-\Delta^2 \sin^2{\theta}
}}=-\frac{I}{2\Delta^2}+\frac{1}{4\Delta}\left(1+\frac{I^2}{\Delta^2}
\right)\ln{\left( \frac{\Delta+I}{\Delta-I} \right)}.
\end{equation}
Integrating now this equation over $I$ we arrive at,
\begin{equation}
K_1(I) =\int\limits_0^I dI \frac{\partial K_1}{\partial
I}=-\frac{I^2}{6\Delta^2}+\frac{I}{4\Delta}\left(1+\frac{I^2}{3\Delta^2}
\right)\ln{\left( \frac{\Delta+I}{\Delta-I} \right)}+\frac{1}{3}
\ln{\left(1-\frac{I^2}{\Delta^2} \right) }.
\end{equation}
Substituting this expression into Eq.~(\ref{gap_planar}) we obtain
the algebraic gap equation,
\begin{equation}
\ln{[1/y]}=-\frac{z^2}{4}+\frac{z}{8}(3+z^2)
\ln{\left(\frac{1+z}{1-z} \right)}+\frac{1}{2} \ln{(1-z^2)},
\end{equation}
where we utilized the same notations as before,
$y=\Delta/\Delta_0$, $z=x/y =I/\Delta$, $x=I/\Delta_0$. The gap
$\Delta_0$ at zero mismatch is $ \Delta_0/2\omega_D=\frac{1}{2}~ \exp(5/6-
{3}/{\nu g})\approx 1.15~\exp{(-3/\nu g)}$. For $x\ll 1$ the
solution to the gap equation has the form, $y=1-{3x^4}/{4}$. Note
that the planar phase is more robust than the polar phase with
respect to surviving the mismatch, in that the gap decreases as the
fourth power for a planar phase instead of the third power for the
polar phase.

Qualitatively, the behavior of the planar phase is very similar to
Fig.~2 with the following numerical values of the characteristic
points: $x_A=0.674$, $y_A=0.787$, $z_A=x_A/y_A=0.856$;
$x_C=e^{-5/6}= 0.435$.

\subsubsection{Large mismatch, $I>\Delta$.}

The gap equation at large mismatch becomes,
\begin{equation}
\label{gap_planar_2}
 -\frac{2}{\nu g}=\int\limits_0^{\pi/2} d\theta \sin^3{\theta}
\ln\left(\frac{I+\sqrt{I^2-\Delta^2 \sin^2{\theta} }}{2\omega_D}\right)\equiv
K_2(I).
\end{equation}
The integral here is calculated by the method used above
[cf.~Eq.~(\ref{der_k1})],
\begin{equation}
\frac{\partial K_2}{\partial I}=\int\limits_0^{\pi/2} d\theta
\frac{\sin^3{\theta}}{ \sqrt{I^2-\Delta^2 \sin^2{\theta}
}}=-\frac{I}{2\Delta^2}+\frac{1}{4\Delta}\left(1+\frac{I^2}{\Delta^2}
\right)\ln{\left( \frac{\Delta+I}{I-\Delta} \right)}.
\end{equation}
Integrating this expression over $I$, and noting that
$K_2(\Delta)=\frac{2}{3}\ln(\Delta/2\omega_D)+\frac{4}{3}\ln{2}-\frac{13}{18}$,
we
obtain,
\begin{equation}
K_2(I) =K_2(\Delta)+\int\limits_\Delta^I dI \frac{\partial
K_2}{\partial I}=\frac{2}{3}
\ln\left(\frac{2\Delta}{2\omega_D}\right)-\frac{5}{9}
-\frac{I^2}{6\Delta^2}+\frac{I}{4\Delta}\left(1+\frac{I^2}{3\Delta^2}
\right)\ln{\left( \frac{\Delta+I}{I-\Delta} \right)}+\frac{1}{3}
\ln{\left(\frac{I^2}{\Delta^2}-1 \right) }.
\end{equation}
We observe that the gap equation in the form
\begin{equation}
\ln{[1/y]}=-\frac{z^2}{4}+\frac{z}{8}(3+z^2)
\ln{\left|\frac{1+z}{1-z} \right|}+\frac{1}{2} \ln{|1-z^2|},
\end{equation}
is actually valid for any relation between $I$ and $\Delta$.
We depict the solutions of the gap equation in the polar
and planar phases in Fig.~2.

\begin{figure}[h]
\label{fig2}
\resizebox{.2\textwidth}{!}{\includegraphics{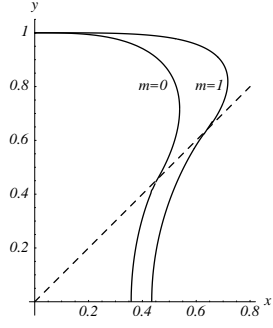}}
\caption{Solutions $y(x)$ of the gap equation in the polar, $m=0$,
and planar, $m=1$, phases. The lower branch corresponds to the
unstable state.   Non-zero solutions of the gap equation cease to
exist beyond the point $y'(x)\rightarrow\infty$, where both
branches merge. The dashed line corresponds to $z=1$.}
\end{figure}

\section{Stability of the obtained solutions}

To analyze the stability of the obtained solutions one has to
calculate the condensation energy for the two phases. The
condensation energy in defined as the difference of the
thermodynamic potentials in the superfluid $\Omega_s$  and
normal $\Omega_n$ states\cite{Abrikosov}.

\subsection{Condensation energy}

The expectation value of the thermodynamic potential at zero
temperature is
\begin{eqnarray}
\Omega_s &=&\sum_{\bf p}\xi_p\left(u_p^2(n_{p,+}+n_{p,-})
+v_p^2(2-n_{p,+}-n_{p,-})\right)+\sum_{\bf p}I_{\Delta}(n_{p,+}-n_{p,-})\nonumber\\
&-&\sum_{{\bf p}{\bf k}}V_{\bf p-k}u_pv_p(1-n_{p,+}-n_{p,-})u_kv_k
(1-n_{k,+}-n_{k,-})
\nonumber\\
&=&\sum_{\bf p}\left(\xi_p-\sqrt{\xi_p^2+\Delta_{\bf p}^2}
(1-n_{p,+}-n_{p,-})\right)\nonumber\\
&+&\sum_{\bf p}\frac{1}{2}\frac{\Delta_{\bf p}^2}{\sqrt{\xi_p^2+\Delta_{\bf p}^2}}
(1-n_{p,+}-n_{p,-})+\sum_{\bf p}I_{\Delta}(n_{p,+}-n_{p,-})
\end{eqnarray}
where we distinguish the mismatches  $I_{\Delta}, I$ respectively with and without
pairing. Here the $u_p$ and $v_p$ are parameters of the
Bogoliubov transformation,
$(u_p^2,v_p^2)=\frac{1}{2}\left(1\pm{\xi_p}/{\sqrt{\xi_p^2+\Delta_{\bf
p}^2}}\right)$, and $n_{p,\pm}$ are the Fermi distributions with
$E_p^{\pm}=\sqrt{\xi_{p}^2+\Delta_{\bf p}^2}\pm I_{\Delta}$.
The difference in the thermodynamic potentials in
the superfluid and normal states is given by,
\begin{eqnarray}
\Omega_s-\Omega_n &=&
\sum_{|\xi_p|>I}|\xi_p|-\sum_{|\xi_p|>\sqrt{I_{\Delta}^2-\Delta_{\bf p}^2}}
\sqrt{\xi_p^2+\Delta_{\bf p}^2}
\nonumber\\
&+&\sum_{|\xi_p|<I}I-\sum_{|\xi_p|<\sqrt{I_{\Delta}^2-\Delta_{\bf p}^2}}I_{\Delta}
+\sum_{|\xi_p|>\sqrt{I_{\Delta}^2-\Delta_{\bf p}^2}}
\frac{1}{2}\frac{\Delta_{\bf p}^2}{\sqrt{\xi_p^2+\Delta_{\bf p}^2}}
\end{eqnarray}
which is readily simplified to
\begin{eqnarray}
\label{condens} \Omega_s-\Omega_n &=&\nu \int \frac{do_{\bf
n}}{4\pi} \left(-\frac{|\Delta_{\bf
n}|^2}{2}-I_{\Delta}\sqrt{I_{\Delta}^2-(\Delta_{\bf n})^2}~
\Theta(I_{\Delta}-|\Delta_{\bf n}|)+I^2 \right).
\end{eqnarray}
Note  that $I=I_{\Delta}$ when pairing occurs at fixed chemical
potentials, but when pairing is considered at fixed numbers
of particles the chemical potentials, in general, are different
before and after pairing: $I\neq I_{\Delta}$.

\subsection{Polar phase}

Applying the formula  (\ref{condens}) to the polar phase at small
mismatch, $z=I/\Delta<1$, and fixed chemical potentials (so
$I_{\Delta}=I$), we obtain
\begin{equation}
\Omega_s-\Omega_n=\nu\Delta^2\left(-\frac{1}{6}-\frac{\pi z^3}{4}
+z^2\right),
\end{equation}
which is negative provided that $z\leq 0.537$. This requires
$y\geq 0.886$. Thus the upper branch $y(x)$ is stable
for $x=I/\Delta_0 \leq 0.475$.  This is the analog of Clogston-Chandrasekhar
limit, here realized through an anisotropic p-wave interaction with the condensate. 
It is easy to
verify that at large mismatches $z=I/\Delta>1$, the condensation
energy
\begin{equation}
\Omega_s-\Omega_n=\nu\Delta^2\left(-\frac{1}{6}-\frac{z^3}{2}
\arcsin\left(z^{-1}\right)-\frac{z}{2}\sqrt{z^2-1} +z^2\right)>0
\end{equation}
is always positive, meaning that the superfluid state is
always unstable.
We plot  in Fig.\ 3 the normalized pressure difference between the
superfluid state with the mismatch $I$ and the normal state
with $I=0$ related to the difference at zero mismatch.
$$P(I)=-\frac{\Omega_s(I)-\Omega_n (0)}{\Omega_s(0)-\Omega_n (0) }.$$
It illustrates that the upper branch is stable for the mismatches
less than the critical mismatch, where the dashed line crosses the
solid line at $x=0.75$. The upper branch is unstable for $x>0.75$.
The lower branch is always unstable.

\begin{figure}[h]
\label{fig3}
\resizebox{.35\textwidth}{!}{\includegraphics{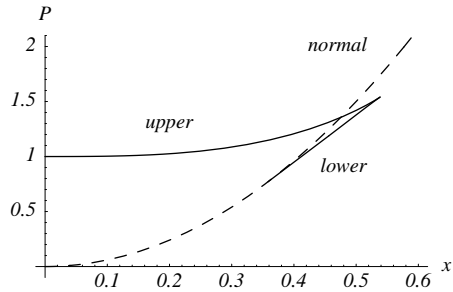}}
\caption{Pressure as a function of the Fermi momenta mismatch
$x=I/\Delta_0$ in the polar (solid line) and normal (broken line)
phases. The upper and lower branches merge at a cusp point
(point A at Fig.1). The lower branch corresponds to the unstable
solution of the gap equation, it is tangent to the normal state at
the point where the superfluid gap disappears $\Delta=0$.
When the pressure of the normal state exceeds the pressure of the
polar phase (upper branch), there is a first order phase
transition at $\Delta\neq 0$ from the superfluid to the
normal state.}
\end{figure}

\subsection{Planar phase}

The condensation energy for the planar phase is easily evaluated
with the help of the integral,
\begin{equation}
\int\limits_0^{\arcsin{a}} d\theta \sin{\theta}
\sqrt{a^2-\sin^2{\theta}} = \frac{a}{2}+\frac{a^2-1}{4}
\ln\left(\frac{1+a}{1-a} \right), ~~~ 0<a<1,
\end{equation}
and yields
\begin{equation}
\Omega_s-\Omega_n=\nu \Delta^2 \left(-\frac{1}{3}
+\frac{z^2}{2}+\frac{z(1-z^2)}{4}
\ln\left|\frac{1+z}{1-z}\right|\right).
\end{equation}
For small mismatches, $z<1$ the energy difference is positive for
$z>0.623$.  For large mismatch $z>1$, the condensation energy is
always positive, $\Omega_s-\Omega_n>0$, indicating that the lower
branch is always unstable. The dependence of the normalized
pressure for the planar phase is similar to Fig.\ 3.

For $I=0$, we obtain the ratio of condensate in the planar phase to the one in polar phase
to be ${\rm e}/2\approx 1.36$, indicating that the planar phase is the ground state
at zero mismatch. This is in agreement with earlier analyses of $l=1$ pairing\cite{Anderson,Tin-LunHo};
the latter contains an extension to higher harmonics.
For our specific model Hamiltonian, at weak coupling, the planar phase
is more stable.

\section{Superfluid density, or ''Meissner'' mass}

In the s-wave version of our model instability associated with spontaneous breakdown
of translation symmetry\cite{WuYip,HuangShov} can arise.   It shows up as a negative superfluid density, or more formally as a negative coefficient 
multiplying the gradient$^2$ terms in the effective Lagrangian for the superfluid mode.   In the context of superconductivity, the gauge invariant form of this term encodes the Meissner mass$^2$ of the photon.  So we can analyze this potential instability by checking whether the Meissner mass$^2$ is positive.  

Following the standard methods in the theory of superconductivity
\cite{AGD} we calculate the super-currents in our system under the
influence of a homogeneous in space gauge field ${\bf A}$, which
is assumed to be transverse. The super-current is anisotropic,
$j_i=\frac{e^2 N}{m}Q_{ik}A_k$ with the components obeying
$Q_{xx}=Q_{yy}$. We consider two cases, when the field ${\bf A}$
is perpendicular and parallel to the direction ${\bf z}$. It is
straightforward to show that the superconducting gap is unaffected
by the external field in the linear response regime.

Let us first consider the case ${\bf A} \cdot {\bf z}=0$.  The
kernel is given by (for $T=0$),
\begin{eqnarray}
\label{kernel} Q_{xx}&=&\lim_{{\bf q} \to 0} \frac{2e^2p_F^2
\nu}{m^2}
\int\limits_{-\infty}^\infty\int\limits_{-\infty}^\infty\frac{d\xi
d\omega}{2\pi}
\frac{do_{\bf n}}{4\pi} \sin^2{\theta} \cos^2{\phi}\nonumber\\
&&\times
\left(\frac{(i\omega+\xi_{p_{+}}-I)(i\omega+\xi_{p_{-}}-I)+\Delta_{p_{+}}\Delta_{p_{-}}}
{((\omega+iI)^2+\xi_{p_{+}}^2+\Delta_{p_{+}}^2)
((\omega+iI)^2+\xi_{p_{-}}^2+\Delta_{p_{-}}^2)}-
\frac{1}{(i\omega-I-\xi_{p_{+}})(i\omega-I-\xi_{p_{-}})}\right),
\end{eqnarray}
where $\theta$ is the angle between ${\bf p}$ and ${\bf z}$, and
$\phi$ is the angle between ${\bf p}$ and the plane containing the
vectors ${\bf z}$ and ${\bf A}$. The density of states $\nu=m
p_F/2\pi^2$ can be taken independent of $I$ (up to $O(I^2/E_F^2)$
corrections). Since the expression (\ref{kernel}) is even in $I$,
the sign of $I$ is irrelevant, so in what follows we take $I>0$
for simplicity. The second term in the parentheses represents the
diamagnetic term (this is easy to verify by calculating first the
integral over the frequency). 
Note, that each term in the parentheses is divergent at large
$\xi$ and $\omega$, but their difference is convergent \cite{AGD}.
Thus, the order of integrals cannot be changed for each term
separately, but it can be changed when calculating the difference
of both terms simultaneously \cite{AGD}.

It is convenient to take first the integral over $d\xi$ and then
the integral over $d\omega$, since the second term in the
parentheses is identically zero when integrating over $d\xi$. In
the remaining term one can immediately assume $q=0$  (the integral
over $d\xi$  commutes with the transition $q\to 0$):
\begin{eqnarray}
\label{kernel1} Q_{xx}=\frac{e^2p_F^2 \nu}{2m^2} \int\limits_0^\pi
{d\theta} \sin^3{\theta} {\cal J}_{\bf n}, ~~~{\cal J}_{\bf n}=
\int\limits_{-\infty}^\infty\frac{d\omega}{2\pi}
\int\limits_{-\infty}^\infty d\xi
\frac{(i\omega+\xi-I)^2+\Delta_{\bf n}^2} {[\xi^2+\Delta^2_{\bf
n}+(\omega+iI)^2]^2}.
\end{eqnarray}
We will now make use of the following identity (which can be
verified by integrating in parts),
\begin{equation}
\label{trans} \int\limits_{-\infty}^\infty d\xi
\frac{(i\omega+\xi-I)^2+\Delta_{\bf n}^2} {[\xi^2+\Delta^2_{\bf
n}+(\omega+iI)^2]^2}=\frac{\Delta^2_{\bf n}}{\Delta^2_{\bf
n}+(\omega+iI)^2} \int\limits_{-\infty}^\infty d\xi \frac{1}
{\xi^2+\Delta^2_{\bf n}+(\omega+iI)^2}.
\end{equation}
The transformation (\ref{trans}) improves the convergence at large
values of $\xi$ and $\omega$. One is now allowed to interchange
the order of the integrals and calculate first the integral over
the frequency with the help of,
\begin{eqnarray}
\label{trans1} \int\limits_{-\infty}^\infty \frac{d\omega}{2\pi}
\frac{1}{[\Delta^2_{\bf n}+(\omega+iI)^2][\xi^2+\Delta^2_{\bf
n}+(\omega+iI)^2]}&=&\frac{1}{2\xi^2} \left(\frac{1}{|\Delta_{\bf
n}|}-\frac{1}{\sqrt{\xi^2+\Delta^2_{\bf n}}}
\right)\Theta\left(|\Delta_{\bf n}|-I\right)\nonumber\\&&
-\frac{1}{2\xi^2 \sqrt{\xi^2+\Delta^2_{\bf
n}}}~\Theta\left(I-|\Delta_{\bf
n}|\right)\Theta\left(\sqrt{\xi^2+\Delta_{\bf n}^2}-I\right).
\end{eqnarray}
Integration over $d\xi$ is now simple and gives,
\begin{eqnarray}
{\cal J}_{\bf n}&=&\Theta \left(|\Delta_{\bf
n}|-I\right)\int\limits_{0}^\infty d\xi\frac{\Delta^2_{\bf
n}}{\xi^2} \left(\frac{1}{|\Delta_{\bf
n}|}-\frac{1}{\sqrt{\xi^2+\Delta^2_{\bf n}}} \right)-
\Theta\left(I-|\Delta_{\bf n}|\right)
\int\limits_{\sqrt{I^2-\Delta_{\bf n}^2}}^\infty
d\xi\frac{\Delta_{\bf n}^2}{\xi^2\sqrt{\xi^2+\Delta^2_{\bf n}}}
\nonumber\\ && =1-\frac{I}{\sqrt{I^2-\Delta_{\bf
n}^2}}~\Theta\left(I-|\Delta_{\bf n}|\right).
\end{eqnarray}
Substituting this expression into Eq.~(\ref{kernel1}) we obtain,
\begin{eqnarray}
\label{kernel2} Q_{xx}&=&\frac{3e^2N}{4m} \int\limits_0^\pi
{d\theta} \sin^3{\theta} \left( 1-\frac{I}{\sqrt{I^2-\Delta_{\bf
n}^2}}~ \Theta\left(I-|\Delta_{\bf n}|\right)\right)
\end{eqnarray}

In the case  of  ${\bf A} \parallel {\bf z}$, 
in complete analogy with the derivation of Eq.~(\ref{kernel2}), we
obtain,
\begin{eqnarray}
\label{kernel3} Q_{zz}&=&\frac{3e^2N}{2m} \int\limits_0^\pi
{d\theta} \sin{\theta} \cos^2{\theta}\left(
1-\frac{I}{\sqrt{I^2-\Delta_{\bf n}^2 }}~
\Theta\left(I-|\Delta_{\bf n}|\right)\right),
\end{eqnarray}
where the integration over the angle $\phi$ gives $1$, instead of
$1/2$ as for the $Q_{xx}$.

Combining both cases, ${\bf A}\perp {\bf z}$ and ${\bf
A}\parallel{\bf z}$, we can now rewrite the general expression for
$Q_{ik}$ as follows,
\begin{equation}
\left( \begin{array}{l}Q_{zz}\\ Q_{xx} \end{array} \right)
=\frac{e^2N}{m}\left(1-\frac{3}{2} \int \frac{do_{\bf n}}{4\pi}
\left(\begin{array}{l}\cos^2{\theta} \\
\sin^2{\theta}\cos^2{\phi}
\end{array} \right)\frac{I}{\sqrt{I^2-\Delta_{\bf n}^2}}~\Theta(I-|\Delta_{\bf n}|)
\right).
\end{equation}
We will now evaluate this expression for the two phases
separately.

\subsection{Polar phase}

The integrals over the angles are easily calculated to yield, for
small mismatches, $z=I/\Delta<1$,
\begin{eqnarray}
Q_{xx} = \frac{Ne^2}{m}\left[1-\frac{3\pi z}{4}+\frac{3\pi
z^3}{8}\right]
\end{eqnarray}
The homogeneous superconducting state is stable provided that this
expression is positive $1-{3\pi} z/{4}+{3\pi z^3}/{8}\geq 0$. This
is so as long as $z\leq 0.480$, which implies that $y> 0.917$ or,
equivalently, $x< 0.440$ (see Fig.4). For larger mismatches, $x>0.440$,
negative values of $Q_{xx}$ probably indicate instability with
respect to the transition into the Larkin-Ovchinnikov-Fulde-Ferrel
state (with paired states of nonzero total momentum).

For large mismatches, $z=I/\Delta>1$, the coefficient
\begin{eqnarray}
Q_{xx} = \frac{Ne^2}{m}\left[1-\frac{3z}{2}
\arcsin\left(z^{-1}\right)+\frac{3z^3}{4}
\arcsin\left(z^{-1}\right)-\frac{3z}{4}\sqrt{z^2-1} \right]
\end{eqnarray}
is always negative,
 indicating instability of the lower branch.

In the case ${\bf A}\parallel{\bf z}$, for small mismatches
$z=I/\Delta<1$, the effective density of superconducting fermions
in the $z$-direction
\begin{eqnarray}
Q_{zz} = \frac{Ne^2}{m}\left[1-\frac{3\pi z^3}{4} \right]
\end{eqnarray}
is positive for $z\leq 0.752$, which implies $y> 0.717$ and $x<
0.589$. This is a weaker condition compared to the one obtained
from the density in the $x$-direction $Q_{xx}$. For large
mismatches $\Delta<I$, the coefficient
\begin{eqnarray}
Q_{zz} = \frac{Ne^2}{m}\left[1-\frac{3z^3}{2}
\arcsin\left(z^{-1}\right)+\frac{3z}{2}\sqrt{z^2-1} \right] <0,
\end{eqnarray}
again demonstrating that the lower branch is unstable.

\begin{figure}[h]
\label{fig3}
\resizebox{.35\textwidth}{!}{\includegraphics{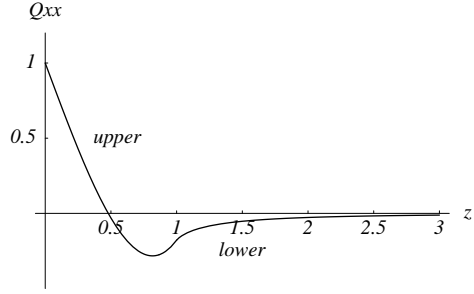}}
\caption{Effective density (``Meissner mass'') of superconducting
fermions in the $x$-direction, $Q_{xx}$,  for the polar phase as
a function of $z=I/\Delta$. The instability sets in at the point
on the upper branch where $Q_{xx}$ becomes negative. The lower
branch
 is always unstable. }
\end{figure}

\subsection{Planar phase}

For the planar phase the calculation of the integrals yields for
both small and large mismatches,
\begin{equation}
\left( \begin{array}{l}Q_{zz}\\ Q_{xx} \end{array}
\right)=\frac{e^2N}{m}\left[1\mp \frac{3z^2}{4}-\frac{3z}{8}\left(
1\mp z^2\right)\ln{\left|\frac{1+z}{1-z} \right|} \right],
\end{equation}
From these expressions we observe that $Q_{xx}$ reaches zero at
$z=0.876$, while $Q_{zz}$ always remains positive. Note that this
critical value exceeds $z_A=0.856$ and $z_D=0.623$ meaning that in
the energetically favorable  state the density of superconducting
fermions is always positive.


\section{Specific heat}
A definitive manifestation of the breached states with gapless
excitations  is the appearance of the term linear in temperature
in the specific heat, which is characteristic for a normal Fermi
liquid. The specific heat is given by the formula,
\begin{equation}
\label{spec_heat}
 C=\sum_{\bf p} \left[E_{\bf p}^{+}\frac{\partial
n(E_{\bf p}^+)}{\partial T}+E_{\bf p}^{-}\frac{\partial n(E_{\bf
p}^-)}{\partial T}\right],
\end{equation}
where $E^{\pm}_{\bf p}=\sqrt{\xi^2_p+\Delta_{\bf n}^2}\pm I$. At
low temperatures $T\ll I$ the first term in Eq.~(\ref{spec_heat})
gives an exponentially small contribution which is negligible. The
second term in Eq.~(\ref{spec_heat}) provides a leading
contribution,
\begin{equation}
\label{spec_heat1} C=\frac{\nu}{4T^2}\int\limits_{-\infty}^\infty
d\xi \int\frac{d \o_{\bf n} }{4\pi}
\frac{\left(\sqrt{\xi^2+|\Delta_{\bf n}|^2}-I
\right)^2}{\cosh^2{\left[\frac{\sqrt{\xi^2+|\Delta_{\bf
n}|^2}-I}{2T}\right]}}.
\end{equation}
We will now evaluate this expression for both phases.

\subsection{Polar phase}

For the polar phase we have,
\begin{equation}
C=\frac{\nu}{4T^2}\int\limits_{-\infty}^\infty
d\xi\int\limits_{-1}^{1}\frac{dx}{2}
\frac{\left(\sqrt{\xi^2+\Delta^2x^2}-I
\right)^2}{\cosh^2{\left[\frac{\sqrt{\xi^2+\Delta^2x^2}-I}{2T}\right]}}.
\end{equation}
At low temperatures $T\ll \Delta$ the integral over the angle $dx$
can be extended to infinity. Rescaling now $x\Delta  = x'$ we
observe that the integral is conveniently rewritten as the
integral over $\rho=\sqrt{\xi_p^2+x^{'^2}}$,
\begin{equation}
C=\frac{\pi \nu}{4T^2\Delta} \int\limits_0^\infty d\rho \rho
\frac{\left(\rho-I
\right)^2}{\cosh^2{\left[\frac{\rho-I}{2T}\right]}}.
\end{equation}
We note that the integrand in this expression is a very sharply
peaked function at $\rho \approx I$ and can, therefore, write,
\begin{equation}
\int\limits_0^\infty d\rho \rho \frac{\left(\rho-I
\right)^2}{\cosh^2{\left[\frac{\rho-I}{2T}\right]}} \to I
\int\limits_0^\infty d\rho \frac{\left(\rho-I
\right)^2}{\cosh^2{\left[\frac{\rho-I}{2T}\right]}} \to  \frac{I
T^3}{2} \int\limits_{-\infty} ^\infty
\frac{dy~y^2}{\cosh^2{[y/2]}}.
\end{equation}
The contribution of the gapless modes to the specific heat is
thus,
\begin{equation}
C=\frac{\pi^2  \nu I }{6\Delta}T,
\end{equation}
which is a fraction $I/4\Delta$ of the specific heat in the normal
state.

\subsection{Planar phase}

The second term in Eq.~(\ref{spec_heat1}) now yields,
\begin{equation}
C=\frac{\nu}{4T^2}\int\limits_{-\infty}^\infty d\xi
\int\limits_{0}^{\pi/2} d\theta \, \sin{\theta }
\frac{\left(\sqrt{\xi^2+\Delta^2 \sin^2{\theta}}-I
\right)^2}{\cosh^2{\left[\frac{\sqrt{\xi^2+\Delta^2\sin^2{\theta}}-I}{2T}\right]}}.
\end{equation}
This integral is a little more tricky for arbitrary ratio between
$I$ and $\Delta$. However, since the principal contribution to the
integral over $d\theta d\xi$ comes from the area where
$\xi_p^2+\Delta^2 \sin^2{\theta}\approx I^2$, we conclude that for
$I\ll \Delta$ the relevant angles are always small $\theta \ll 1$,
\begin{equation}
C=\frac{\nu}{4T^2}\int\limits_{-\infty}^\infty d\xi
\int\limits_{0}^{\infty} d\theta \, \theta
\frac{\left(\sqrt{\xi^2+\Delta^2 \theta^2}-I
\right)^2}{\cosh^2{\left[\frac{\sqrt{\xi^2+\Delta^2
\theta^2}-I}{2T}\right]}},
\end{equation}
where again we can extend the limits of the angle integration to
infinity due to fast convergence ($\Delta \gg T$) of the integral.
Transforming this integral to the polar coordinates similar to the
above ($\rho=\sqrt{\xi^2+\Delta^2\theta^2}$) one obtains,
\begin{equation}
C=\frac{\nu}{2T^2\Delta} \int\limits_0^\infty d\rho \rho^2
\frac{\left(\rho-I
\right)^2}{\cosh^2{\left[\frac{\rho-I}{2T}\right]}} \to \frac{\nu
I^2 }{2T^2\Delta^2} \int\limits_{-\infty}^\infty d\rho
\frac{\left(\rho-I
\right)^2}{\cosh^2{\left[\frac{\rho-I}{2T}\right]}} =
\frac{2\pi}{3}\frac{\nu I^2}{\Delta^2} T.
\end{equation}

This result is valid provided that $T\ll I\ll \Delta$. As might be
expected, the linear in $T$ contribution to the specific heat is
proportional to the area occupied by the gapless modes, i.e. the
$\sim I/\Delta$ strip around the equator for the polar phase and
the islands around the poles for the planar phase.

\section{Competition of the anisotropic and mixed phases}

In this section we analyze the stability of the obtained phases
with respect to the transition into a spatially inhomogeneous
state under the condition of {\it fixed number of particles\/}
rather than at fixed chemical potentials as assumed in the
preceding sections. Competition arises between a phase with
the finite Fermi momentum mismatch $I$ (i.e.\ with unequal numbers
of particles) and a mixed phase consisting of spatially separated
regions of superfluid state with zero mismatch (equal number
of particles)  and a normal (unpaired) phase accommodating the
extra particles. To analyze the stability under the condition of a
fixed number of particles one has to evaluate the energy (rather
than the thermodynamic potential) of both phases.

\subsection{Mixed phase}

The energy of the mixed phase is given by the sum of energies of the
superfluid and normal states, see Ref. (6) for the s wave,
\begin{eqnarray}
E_{mix} &=& (1-x)\left(\frac{(6\pi^2\bar{n})^{4/3}\,v}{4\pi^2}
-\frac{(6\pi^2\bar{n})^{2/3}}{2\pi^2\,v}\frac{\Delta_0^2}{2c}
\right)
+ x\left(\frac{(6\pi^2\bar{n}_A)^{4/3}\,v}{8\pi^2}
+\frac{(6\pi^2\bar{n}_B)^{4/3}\,v}{8\pi^2}
\right)
\end{eqnarray}
where $x/1-x$ fraction of the volume is in the
normal/superfluid state (thermodynamic limit), total numbers
of particles (densities) are $n_A=x\bar{n}_A+(1-x)\bar{n}$ and
$n_B=x\bar{n}_B+(1-x)\bar{n}$, the superfluid state is given
by the anisotropic pairing at zero mismatch containing $\bar{n}$
particles of each species. We introduced the constant $c$ which
assumes values $c=3$ for the polar phase and $c=3/2$ for the
planar phase, respectively. There are two variational parameters,
$x$ and $\bar{n}$. It is convenient to introduce the new variables
$\bar{n}=n_{A}+\delta\bar{n}$ and $\delta n=n_B-n_A$. At small
mismatches we obtain up to the second order
\begin{eqnarray}
E_{mix} &=& \frac{(6\pi^2n_A)^{4/3}\,v}{4\pi^2}
\left(1+\frac{2}{3}\frac{\delta n}{n_A}+\frac{1}{9x}\frac{\delta n^2}{n_A^2}
+\frac{2}{9}\frac{1-x}{x}\frac{\delta\bar{n}^2}{n_A^2}
-\frac{2}{9}\frac{1-x}{x}\frac{\delta n\delta\bar{n}}{n_A^2}
\right)-(1-x)\frac{(6\pi^2n_A)^{2/3}}{2\pi^2\,v}\frac{\Delta_0^2}{2c}
\end{eqnarray}
Subtracting the energy of the normal state,
\begin{eqnarray}
E_{n} &=& \frac{(6\pi^2 n_A)^{4/3}\,v}{8\pi^2}
+\frac{(6\pi^2 n_B)^{4/3}\,v}{8\pi^2}
=\frac{(6\pi^2n_A)^{4/3}\,v}{4\pi^2}\left(1+\frac{2}{3}\frac{\delta n}{n_A}
+\frac{1}{9}\frac{\delta n^2}{n_A^2}
\right)
\end{eqnarray}
and minimizing $E_{mix}-E_n$ with respect to $\delta\bar{n}$ and $x$,
we obtain
\begin{eqnarray}
\label{emix} E_{mix}-E_n &=&
-\frac{(6\pi^2n_A)^{2/3}}{2\pi^2\,v}\frac{\Delta_0^2}{2c}
(1-x_{min})^2 \nonumber\\
x_{min} &=& \frac{(6\pi^2n_A)^{1/3}\,v\sqrt{c}}{3\sqrt{2}\Delta_0}\frac{\delta
n}{n_A}
\end{eqnarray}
where $\Delta_0$ is different for the polar, $\Delta_0^{pol}$,
and the planar, $\Delta_0^{pl}$, phases.

\subsection{Energy of the anisotropic phases}

It is convenient to start from a thermodynamic potential at fixed
chemical potentials and to perform a Legendre transformation to
find the energy expressed in terms of fixed particle numbers. We
begin by calculating the thermodynamic potential of the
mismatched superfluid state,
\begin{eqnarray}
\Omega_s (I_\Delta)-\Omega_0 &=& \sum_{\bf p}|\xi_{\bf
p}|-\sum_{|\xi_{\bf p}|>\sqrt{I_{\Delta}^2-\Delta_{\bf p}^2}}
\sqrt{\xi_{\bf p}^2+\Delta_{\bf p}^2}-\sum_{|\xi_{\bf
p}|<\sqrt{I_{\Delta}^2-\Delta_{\bf p}^2}}I_{\Delta}
+\frac{1}{2}\sum_{|\xi_p|>\sqrt{I^2-\Delta_{\bf
p}^2}}\frac{\Delta_p^2} {\sqrt{\xi_{\bf p}^2+\Delta_{\bf p}^2}},
\end{eqnarray}
counted from the potential of the normal state at zero mismatch,
$\Omega_0 =\sum_{\bf p}\xi_{\bf p}-\sum_{\bf p}|\xi_{\bf p}|$; and
we use the notation $I_{\Delta}$ for the mismatch in the
superfluid state.  The calculation similar to that from
Section III gives,
\begin{eqnarray}
\Omega_s(I_{\Delta})=\Omega_0+\nu\int\frac{do_{\bf n}}{4\pi}
\left(-\frac{\Delta_{\bf
n}^2}{2}-I_{\Delta}\sqrt{I_{\Delta}^2-\Delta_{\bf n}^2} ~
\Theta(I-|\Delta_{\bf n}|)\right).
\end{eqnarray}
We can now perform the Legendre transformation to find the energy
of the superfluid state,
\begin{eqnarray}
E_s&=&\Omega_s+\frac{\mu_A+\mu_B}{2}(n_A+n_B)+\frac{\mu_A-\mu_B}{2}(n_A-n_B)
\nonumber\\
&=& \Omega_0+p_F(n_A+n_B) +\nu\int\frac{do_{\bf n}}{4\pi}
\left(-\frac{\Delta_{\bf
n}^2}{2}-I_{\Delta}\sqrt{I_{\Delta}^2-\Delta_{\bf n}^2}\right)
+I_{\Delta}\delta n,
\end{eqnarray}
here $\delta n=n_B-n_A$ is the difference in the number of
particles corresponding to the chemical potential mismatch
$I_\Delta$,
\begin{eqnarray}
\label{delta n} \delta n&=& \sum_{|\xi_{\bf
p}|<\sqrt{I_{\Delta}^2-\Delta_{\bf p}^2}}
\Theta(I_{\Delta}-\sqrt{\xi_{\bf p}^2+\Delta_{\bf p}^2}) =
2\nu\int\frac{do_{\bf n}}{4\pi}\sqrt{I_{\Delta}^2-\Delta_{\bf
n}^2}\Theta(I-|\Delta_{\bf n}|).
\end{eqnarray}
The condensation energy (the difference in the energy of a
superfluid state and a normal state) can now be written as,
\begin{eqnarray}
\label{energy_diff} E_s-E_n&=& \nu\int\frac{do_{\bf n}}{4\pi}
\left(-\frac{\Delta_{\bf
n}^2}{2}+I_{\Delta}\sqrt{I_{\Delta}^2-\Delta_{\bf n}^2}-I^2\right)
\end{eqnarray}
where we introduced the chemical potential mismatch $I=\delta
n/2\nu$ in the normal state, corresponding to the particle number
mismatch $\delta n$.
Note that in deriving Eq.~(\ref{energy_diff}) we also used that
at non-zero mismatch, $E_n=E_0+\nu I^2$ and $\Omega_n=\Omega_0-\nu
I^2$ in the normal state.

\subsection{Polar phase}
Applying the above expression (\ref{energy_diff}) for the polar
phase, $\Delta_{\bf n}=\Delta\cos\theta$,
\begin{eqnarray}
E_s-E_0
&=&\nu\left(-\frac{\Delta^2}{6}+\frac{I_{\Delta}^3}{\Delta}\frac{\pi}{4}\right),
\end{eqnarray}
where $E_0$ is the energy of the normal state with no mismatch,
and the chemical potential mismatch $I_\Delta$ is related to the
particle number difference $\delta n$ as,
\begin{equation} \delta n = \frac{\pi \nu I_{\Delta}^2 }{2\Delta}.
\end{equation}
We are interested in the case of small mismatches, $I_\Delta \ll
\Delta_0$, where it is possible to approximate,
${\Delta}/{\Delta_0} = 1-\pi I_{\Delta}^4/4 \Delta_0^4$. Using
these expressions we obtain to the lowest non-vanishing order,
\begin{eqnarray}
E_s-E_0=-\nu\frac{\Delta_0^2}{6}\left[1-4\sqrt{\frac{2}{\pi}}
\left(\frac{\delta n}{\nu\Delta_0}\right)^{3/2}
\right]
\end{eqnarray}
Comparing it with the energy (\ref{emix}) of the mixed phase
(which is simplified using, $\nu=(6\pi^2 n_A)^{2/3}/(2\pi^2\,v)$ and
$x_{min}=\sqrt{3/2}~(\delta n/\nu \Delta_0)$),
\begin{eqnarray}
\label{em-e0} E_{mix}-E_0=-\nu\frac{\Delta_0^2}{6}\left[1-\sqrt{6}
\left(\frac{\delta n}{\nu\Delta_0}\right)
\right]
\end{eqnarray}
we see that the homogeneous superfluid state is
energetically favorable for small mismatches $\delta n \ll \nu
\Delta_0$.

\subsection{Planar phase}

The energy of the planar phase, $\Delta_{\bf
n}=\Delta\sin\theta{\rm e}^{\pm i\phi}$, counted from the energy
of the normal state with zero mismatch is
\begin{eqnarray}
\label{es-planar}
 E_s-E_0
&=&\nu\left(-\frac{\Delta^2}{3}+\frac{I_{\Delta}^2}{2}
-\frac{\Delta
I_{\Delta}}{4}\left(1-\frac{I_{\Delta}^2}{\Delta^2}\right)
\ln\left[\frac{1+I_{\Delta}/\Delta}{1-I_{\Delta}/\Delta}\right]
\right).
\end{eqnarray}
The particle difference is related to the chemical potential
mismatch according to,
\begin{eqnarray}
\label{deltanplanar} \delta
n=\nu\left(I_{\Delta}-\frac{\Delta}{2}\left(1-\frac{I_{\Delta}^2}{\Delta^2}\right)
\ln\left[\frac{1+I_{\Delta}/\Delta}{1-I_{\Delta}/\Delta}\right]\right).
\end{eqnarray}
For small mismatches $\delta n\ll \nu \Delta_0$, one can
approximate, $\delta n ={2\nu I_\Delta^3}/{3\Delta^2}$. The gap at
small mismatches has the form, ${\Delta}/{\Delta_0} = 1-{3
I_{\Delta}^4}/4{\Delta^4}$. Substituting this expression into
Eq.~(\ref{es-planar}) and eliminating $\delta n$ with the help of
Eq.~(\ref{deltanplanar}) we obtain,
\begin{eqnarray}
E_s-E_0 &=& -\nu\frac{\Delta_0^2}{3}\left[1-5\frac{ 3^{4/3}}{
2^{7/3}} \left(\frac{\delta n}{\nu\Delta_0}\right)^{4/3} \right]
\end{eqnarray}
Again, comparing this expression to the energy (\ref{emix}) of the
mixed phase (with $x_{min}=\sqrt{3}/2(\delta n/\nu\Delta_0)$),
\begin{eqnarray}
E_{mix}-E_0 &=& -\nu\frac{\Delta_0^2}{3}\left[1-\sqrt{3}
\left(\frac{\delta n}{\nu\Delta_0}\right) \right],
\end{eqnarray}
we conclude that the superfluid state is more stable than the mixed state.

\section{Conclusion}

We have presented substantial theoretical evidence that our simple model
supports planar phase gapless superfluidity in the ground
state.  For  $I\ll \Delta$ the gapless modes contribute terms of high
order in the mismatch, $\sim I^4$ in the solution of the gap equation and $\sim
I^2$ in the heat capacity, i.e. they represent small
perturbations.   The planar phase is symmetric under simultaneous axial
rotation and gauge (i.e., phase) transformation. The residual continuous symmetry of this state,
and its favorable energy relative to plausible competitors (normal
state, polar phase) suggest that it is a true ground state in this
model.  Also, we obtain a
positive density of superfluid fermions, suggesting that
inhomogeneous LOFF phases are disfavored at small $I$. Direct
calculation of the energies in the anisotropic superfluid and
mixed phases shows that the p-wave breached superfluidity is
energetically favorable and the phase separation does not occur at
small mismatch.

In some respects the same qualitative behavior we find here in the
p wave resembles what arose in s wave \cite{FGLW}. Namely,
isotropic s-wave superfluidity has two branches of solution:
the upper BCS which is stable and -- for simple interactions --
fully gapped, and the lower branch which has gapless modes but is
unstable. The striking difference is that in p wave the upper
branch retains its stability while it develops two-dimensional lenticular surfaces
of gapless modes.  Specifically, the anisotropic p-wave breached pair
phase, with coexisting superfluid and normal components, is stable
already for a wide range of parameters at weak coupling using the
simplest (momentum-independent) interaction.  This bodes well for
its future experimental realization.

In our model, which has no explicit spin degree of freedom,
gapless modes occur for either choice of order parameter with
residual continuous symmetry. By contrast, for $^{3}$He in the B
phase the p-wave spin-triplet order parameter is a $2\times 2$
spin matrix, containing both polar and planar phases components,
there are no zeros in the quasiparticle energies, and the
phenomenology broadly resembles that of a conventional s-wave
state \cite{BW}; in the A phase (which arises only at  $T\neq 0$
\cite{Leggett}) the separate up and down spin components pair with
themselves, in an orbital p wave, and no possibility of a mismatch
arises.

Application of anisotropic breached superfluidity to high density
QCD may lead to a viable stable phase at low chemical potential.
Though the gap with higher orbital harmonics is suppressed, the
smallest gap defines neutrino emission properties and hence
cooling rate of the neutron star\cite{neutronStar}.

It is possible that the emergent fermi gas of gapless excitations
develops, as a result of residual interactions, secondary
condensations. Also, one may consider analogous possibilities for
particle-hole, as opposed to particle-particle, pairing. In that
context, deviations from nesting play the role that fermi surface
mismatch plays in the particle-particle case. We are actively
investigating these issues.

\section*{Acknowledgements}

The authors thank E. Demler, M. Forbes, O. Jahn, R. Jaffe, B.
Halperin, V. Liu, G. Nardulli, A. Scardicchio, O. Schroeder, A. Shytov, I.
Shovkovy, D. Son for useful discussions. This work is
supported in part by funds provided by the U. S. Department of
Energy (D.O.E.) under cooperative research agreement
DF-FC02-94ER40818, and by DOE grant DE-FG02-06ER46313.


\begin{thebibliography}{99}


\bibitem{Sarma} G. Sarma, Phys. Chem. Solid {\bf 24}, 1029 (1963); S. Takada and
T. Izuyama, Prog. Theor. Phys. {\bf 41}, 635 (1969).

\bibitem{LW} V. Liu and F. Wilczek,
{\it Phys. Rev. Lett.} {\bf 90}, 047002 (2003).

\bibitem{GLW} E. Gubankova, V. Liu, and F. Wilczek,
{\it Phys. Rev. Lett.} {\bf 91}, 032001 (2003).

\bibitem{ABR} M. Alford, J. Berges, and K. Rajagopal,
{\it Phys.Rev.Lett.} {\bf 84}, 598 (2000).

\bibitem{SH} I. Schovkovy and M. Huang,
{\it Phys. Lett. B} {\bf 564}, 205 (2003).

\bibitem{Bedaque} P. F. Bedaque, H. Caldas, and G. Rupak,
{\it Phys. Rev. Lett.} {\bf 91}, 247002 (2003).

\bibitem{WuYip} S. T. Wu and S. Yip,
{\it Phys. Rev. A} {\bf 67}, 053603 (2003).

\bibitem{HuangShov} M. Huang and I. Shovkovy,
{\it Phys. Rev. D} {\bf 70}, 051501 (2004).

\bibitem{Nardulli} R. Casalbuoni, R. Gatto, M. Mannarelli, G. Nardulli, M. Ruggieri,
{\it Phys. Lett. B} {\bf 605}, 362 (2005);
{\bf 615}, 297(E) (2005).

\bibitem{FGLW} M. Forbes, E. Gubankova, V. Liu, and F. Wilczek,
{\it Phys. Rev. Lett.} {\bf 94}, 017001 (2005).

\bibitem{AKR} M. Alford,  C. Kouvaris, and K. Rajagopal,
hep-ph/0407257.

\bibitem{coldatoms} For a review, see Nature 416, 205 (2002).

\bibitem{exp} G. Modugno, et.al., Science, {\bf 297}, 2240 (2002);
J. Goldwin, et.al., {\it Phys. Rev. A} {\bf 70}, 021601 (2004).

\bibitem{Zhang} J. Zhang, et.al., quant-ph/0406085.

\bibitem{Son} D. T. Son and M. A. Stephanov,
{\it Phys. Rev. A} {\bf 74}, 013614 (2006);
C. H. Pao, S. T. Wu, and S. K. Yip,
{\it Phys. Rev. B} {\bf 73}, 132506 (2006);
{\bf 73}, 132506(E) (2006).

\bibitem{recent} E. Gubankova, A. Schmitt, and F. Wilczek,
{\it Phys. Rev. B} {\bf 74}, 064505 (2006).

\bibitem{Kohn-Luttinger} T. Schaefer,
{\it Phys. Rev. D} {\bf74}, 054009 (2006).

\bibitem{Schaefer} T. Schaefer,
{\it Phys. Rev. D} {\bf62}, 094007 (2000).

\bibitem{shortVersion}E. Gubankova, E. Mishchenko, and F. Wilczek,
 {\it Phys. Rev. Lett.} {\bf 94}, 110402 (2005).

\bibitem{Mineev} V. P. Mineev and K. V. Samokhin,
{\it Introduction to Unconventional Superconductivity}, (Gordon
and Breach Science Publishers, 1999).

\bibitem{Abrikosov} A. A. Abrikosov, {\it Fundamentals of the Theory of Metals},
(Elsevier, 1988).

\bibitem{LOFF} A. I. Larkin and Yu. N. Ovchinnikov,
{\it Sov. Phys. JETP} {\bf 20}, 762 (1965); P. Fulde and R. A.
Ferrell, {\it Phys. Rev.} {\bf 135}, A550 (1964).

\bibitem{yangSondhi}K. Yang and S. Sondhi, {\it Phys. Rev}. {\bf B57}, 8566 (1998).

\bibitem{Volovik} G. E. Volovik,
{\it Phys. Lett. A} {\bf 142}, 282 (1989).

\bibitem{BW} R. Balian and N. R. Werthamer,
{\it Phys. Rev.} {\bf 131}, 1553 (1963).

\bibitem{Anderson} P. W. Anderson,
{\it Phys. Rev.} {\bf 112}, 1900 (1958).

\bibitem{Tin-LunHo} T.-L. Ho and R. B. Diener,
{\it Phys. Rev. Lett.} {\bf 94}, 090402 (2005).

\bibitem{AGD} A. A. Abrikosov, L. P. Gorkov, and I. E.
Dzyaloshinskii, {\it Methods of Quantum Field Theory in
Statistical Physics}, (Dover, New York, 1975).

\bibitem{Leggett} A. J. Leggett, Rev. Mod. Phys. {\bf 47}, 331 (1975).

\bibitem{neutronStar} M. G. Alford, J. A. Bowers, J. M. Cheyne, G. A. Cowan,
{\it Phys. Rev. D} {\bf 67}, 054018 (2003).

\end{thebibliography}
\end{document}